\documentstyle[aps,floats,eqsecnum,psfig]{revtex}

\begin{document}
\draft

\title{Scalar GW detection with a hollow spherical antenna}

\author{E.Coccia, F. Fucito and M.Salvino}
\address{Dipartimento di Fisica, Universit\`a di Roma ``Tor Vergata''
	{\rm and\/} \\ INFN Sezione di Roma Tor Vergata, Via Ricerca 
	Scientifica 1, 00133 Roma, Italy}
\author{J. A. Lobo}
\address{Departament de F\'\i sica Fonamental, Universitat de Barcelona,
	Spain}

\maketitle

\begin{abstract}
We study the response and cross sections for the absorption of GW energy
in a Jordan-Brans-Dicke theory by a resonant mass detector shaped
as a hollow sphere.

\end{abstract}

\section{Introduction}
It seems reasonable to predict that the new gravitational wave (GW) detectors
now under construction, once operating at the maximum of their sensitivity,
will be able to detect GWs. Will it be possible to use these future
measurements to try to gain information on which is the theory of gravity
at low energies? There are no particular reasons, in fact, why GW must be
of spin two. In reality, many theories of gravity can be built which contain
scalars and vectors. These theories are mathematically well founded.
String theory, in particular, is believed to be consistent also as
a quantum description of gravity. The predictions of these theories must
then be checked against available experimental data. This forces the
couplings and masses present in the Lagrangian to take values in well
defined domains. See \cite{will} for a more detailed exposition. Once
detected, one can also attempt to use GWs as a means to further constrain
this picture. It seems relevant to try to develop the theory to the
point where it can profit from new experimental insights. For these reasons,
it has been analysed in great detail in reference \cite{bian-98} ---see also
\cite{lobo-95,wp-75} the interaction and cross section of an elastic massive
sphere with scalar waves. 

An appealing variant of the massive sphere is a {\it hollow\/} sphere
\cite{vega-98}. The latter has the remarkable property that it enables the
detector to monitor GW signals in a significantly {\it lower frequency range\/}
---down to about 200 Hz--- than its massive counterpart for comparable
sphere masses. This can be considered a positive advantage for a future
world wide network of GW detectors, as the sensitivity range of such
antenna overlaps with that of the large scale interferometers, now in
a rather advanced state of construction \cite{ligo,virgo}.
Moreover it appears technically easier to fabricate in large 
dimensions \cite{vega-98}.  

A hollow sphere obviously has the same symmetry of the massive one, so
the general structure of its {\it normal modes\/} of vibration is very
similar in both \cite{vega-98}. In particular, the hollow sphere is
very well adapted to sense and monitor the presence of scalar modes in
the incoming GW signal. In this paper we extend the analysis of the response 
of a hollow sphere \cite{vega-98}, to include scalar excitations. 

In section II we briefly review the normal mode
algebra of the hollow sphere; then in section III we calculate the
scalar cross sections for the absorption of GW energy in scalar modes,
and in section IV we assess the detectability of a few interesting
sources on the assumption that they behave as Jordan-Brans-Dicke
emitters \cite{jordan-59,bd-61} of GWs. Finally, section V is devoted
to a summary of conclusions.

\section{Review of hollow sphere normal modes}

This section contains some review material which is included essentially to
fix the notation and to ease the reading of the ensuing sections. 
The eigenmode equation for a
three-dimensional elastic solid is the following:

\begin{equation}
\nabla^2{\bf s}+\left(1+\lambda/\mu\right)
\nabla (\nabla\cdot{\bf s})=-k^2{\bf s}\ ,\qquad
\left(k^2\equiv\varrho\omega^2/\mu\right),
\label{1.1}
\end{equation}
as described in standard textbooks, such as \cite{love-44,landau-70}. 
$\lambda$ and $\mu$ are the material's Lam\'e coefficients, $\rho$ is the
material density and $\omega$ is the angular frequency.

The equation must be solved subject to the {\it boundary conditions} that the
solid is to be free from tensions and/or tractions. In the case of a hollow
sphere, we have two boundaries given by
the outer and the inner surfaces
of the solid itself. We use the notation $a\/$ for the inner 
radius, and $R\/$ for
the outer radius. The boundary conditions are thus expressed by

\begin{equation}
\sigma_{ij}n_j=0\hspace{1cm}\mbox{at}\hspace{0.4cm} r=R
\hspace{0.4cm}\mbox{and at}\hspace{0.4cm}r=a\hspace{0.5cm}(R\geq a \geq 0),
\label{bc}
\end{equation}
where $\sigma_{ij}\/$ is the stress tensor, and is given by \cite{landau-70}

\begin{equation}
\sigma_{ij}=\lambda\,u_{k,k}\, \delta_{ij}\,+\,2\,\mu \,u_{(i,j)}.            
\end{equation}
where {\bf n} is
the unit, outward pointing normal vector.

The general solution to equation (\ref{1.1}) is a linear superposition of
a longitudinal vector field and two transverse vector fields, i.e.,

\begin{equation}
{\bf s}(r,\vartheta,\phi) =
\frac{C_{\rm l}}{q}\,{\bf s}_{\rm l} + \frac{C_{\rm t}}{k}\,{\bf s}_{\rm t}
+ C_{\rm t'}\,{\bf s}_{\rm t'}
\label{1.25}
\end{equation}
where $C_{\rm l}$, $C_{\rm t}$ and $C_{\rm t'}$ are constant coefficients,
and

\begin{mathletters}
\label{1.3}
\begin{eqnarray}
{\bf s}_{\rm l}\left(r,\vartheta,\phi \right) &=& \frac{dh_l(qr,E)}{dr}\,
Y_{lm}{\bf n} - \frac{h_l(qr,E)}{r}\,i{\bf n}\times {\bf L}Y_{lm}
\label{1.3a} \\
{\bf s}_{\rm t}\left(r,\vartheta,\phi\right) &=& -l\left(l+1\right)
\frac{h_l\left(kr,F\right)}{r}\,Y_{lm}{\bf n} + \left[
\frac{h_l(kr,F)}{r} + \frac{dh_l\left(kr,F\right)}{dr}\right)]\,
i{\bf n}\times {\bf L}Y_{lm} \label{1.3b} \\
{\bf s}_{\rm t'}\left(r,\vartheta,\phi\right) &=&
h_l(kr,F)\,i{\bf L}Y_{lm} \label{1.3c}
\end{eqnarray}
\end{mathletters}
with $E\/$ and $F\/$ also arbitrary constants,

\begin{equation}
q^2\equiv k^2\,\frac{\mu}{\lambda+\mu} =
\frac{\varrho_0\omega^2}{\lambda+\mu}
\label{1.35}
\end{equation}
and

\begin{equation}
h_l\left(z,A\right)\equiv j_l(z) + A\,y_l(z)
\label{1.4}
\end{equation}
$j_l\/$ $y_l\/$ are spherical Bessel functions \cite{abram-72}:

\begin{mathletters}
\label{1.5}
\begin{eqnarray}
j_l(z) &=& z^l\,\left(-\frac 1z\,\frac d{dz}\right)^l\,\frac{\sin z}z
\label{1.5a} \\
y_l(z) &=& -z^l\,\left(-\frac 1z\,\frac d{dz}\right)^l\,\frac{\cos z}z
\label{1.5b}
\end{eqnarray}
\end{mathletters}

Finally, {\bf L\/} is the {\it angular momentum\/} operator

\begin{equation}
{\bf L}\equiv -i\,{\bf x}\times\nabla
\label{1.6}
\end{equation}

The boundary conditions (\ref{bc}) must now be imposed on the generic solution
to equations (\ref{1.1}). After some rather heavy algebra it is finally found
that there are two families of eigenmodes, the {\it toroidal\/} (purely
rotational) and the {\it spheroidal\/}. Only the latter couple to GWs
\cite{bian-96}, so we shall be interested exclusively in them. The form
of the associated wavefunctions is

\begin{equation}
{\bf s}_{nlm}^S(r,\vartheta ,\phi) =
A_{nl}(r)\,Y_{lm}(\vartheta ,\phi)\,{\bf n} -
B_{nl}(r)\,i{\bf n}\times{\bf L}Y_{lm}(\vartheta ,\phi)
\label{1.7}
\end{equation}
where the radial functions $A_{nl}(r)$ and $B_{nl}(r)$ have rather
complicated expressions:

\begin{mathletters}
\label{1.8}
\begin{eqnarray}
A_{nl}(r) &=& C(nl)\,\left[\frac{1}{q_{nl}^S}\,\frac{d}{dr}\,
j_l(q_{nl}^Sr) -
l(l+1)\,K(nl)\,\frac{j_l(k_{nl}^Sr)}{k_{nl}^Sr}+\right.
  \nonumber \\
&& \ \qquad\quad + \left. D(nl)\,\frac{1}{q_{nl}^S}\,\frac{d}{dr}\,
y_l(q_{nl}^Sr)
- l(l+1)\,\tilde D(nl)\,\frac{y_l(k_{nl}^Sr)}{k_{nl}^Sr}\right]
\label{1.8a}  \\
B_{nl}(r) &=& C(nl)\,\left[\frac{j_l(q_{nl}^Sr)}{q_{nl}^Sr} -
K(nl)\,\frac 1{k_{nl}^Sr}\,\frac d{dr}\left\{r\,j_l(k_{nl}^Sr)\right\} +
\right.  \nonumber \\
&& \ \qquad\quad + \left. D(nl)\,\frac{y_l(q_{nl}^Sr)}{q_{nl}^Sr} -
\tilde D(nl)\,\frac 1{k_{nl}^Sr}\,\frac d{dr}\left\{r\,y_l(k_{nl}^Sr)\right\}
\right] \label{1.8b}
\end{eqnarray}
\end{mathletters}

Here $k_{nl}^SR$ and $q_{nl}^SR$ are dimensionless {\it eigenvalues\/},
and they are the solution to a rather complicated algebraic equation for
the frequencies $\omega\/$\,=\,$\omega_{nl}\/$ in (\ref{1.1}) ---see
\cite{vega-98} for details. In (\ref{1.8a}) and (\ref{1.8b}) we have set

\begin{equation}
K(nl)\equiv\frac{C_{\rm t}q_{nl}^S}{C_{\rm l}k_{nl}^S}\ ,\qquad
D(nl)\equiv\frac{q_{nl}^S}{k_{nl}^S}\,E\ ,\qquad
\tilde D(nl)\equiv\frac{C_{\rm t}Fq_{nl}^S}{C_{\rm l}k_{nl}^S}
\label{1.9}
\end{equation}
and introduced the normalisation constant $C(nl)$, which is fixed by the
orthogonality properties

\begin{equation}
\int_V({\bf s}_{n^{\prime}l^{\prime}m^{\prime}}^S)^*\cdot
({\bf s}_{nlm}^S)\,\varrho_0\,d^3 x =
M\,\delta_{nn^{\prime}}\delta_{ll^{\prime}}\delta_{mm^{\prime}} 
\label{1.10}
\end{equation}
where $M\/$ is the mass of the hollow sphere:

\begin{equation}
M = \frac{4\pi}3\,\varrho_0 R^3\,(1-\varsigma ^3)\ ,\qquad
\varsigma\equiv\frac{a}{R}\leq 1
\label{1.11}
\end{equation}

Equation (\ref{1.10}) fixes the value of $C(nl)$ through the radial integral

\begin{equation}
\int_{\varsigma R}^R\,\left[A_{nl}^2(r) + l(l+1)\,B_{nl}^2(r)\right]\,
r^2dr = \frac{4\pi}3\varrho_0\,(1-\varsigma^3)R^3
\label{1.12}
\end{equation}
as can be easily verified by suitable manipulation of (\ref{1.7}) and the
well known properties of angular momentum operators and spherical harmonics.
We shall later specify the values of the different parameters appearing in
the above expressions as required in each particular case which will in due
course be considered.

\section{Absorption cross sections}

As seen in reference \cite{lobo-95}, a scalar--tensor theory of GWs such as
JBD predicts the excitation of the sphere's monopole modes {\it as well as
the\/} $m\/$\,=\,0 quadrupole modes. In order to calculate the energy absorbed
by the detector according to that theory it is necessary to calculate the
energy deposited by the wave in those modes, and this in turn requires that
we solve the elasticity equation with the GW driving term included in its
right hand side. The result of such calculation was presented in full
generality in reference \cite{lobo-95}, and is directly applicable here
because the structure of the oscillation eigenmodes of a hollow sphere is
equal to that of the massive sphere ---only the explicit form of the
wavefunctions needs to be changed. We thus have

\begin{equation}
E_{\rm osc}(\omega_{nl}) = \frac{1}{2}\,Mb^2_{nl}
  \,\sum_{m=-l}^l\,|G^{(lm)}(\omega_{nl})|^2
\label{2.0}
\end{equation}
where $G^{(lm)}(\omega_{nl})$ is the Fourier amplitude of the corresponding
incoming GW mode, and

\begin{mathletters}
\label{2.1}
\begin{eqnarray}
b_{n0} &=& -\frac{\varrho_0}{M}\,\int_a^R\,A_{n0}(r)\,r^3 dr
\label{2.1a} \\[1 ex]
b_{n2} &=& -\frac{\varrho_0}{M}\,\int_a^R\,\left[A_{n2}(r)
	 + 3B_{n2}(r)\right]\,r^3 dr
\label{2.1b}
\end{eqnarray}
\end{mathletters}
for monopole and quadrupole modes, respectively, and $A_{nl}(r)$ and
$B_{nl}(r)$ are given by (\ref{1.8}). Explicit calculation yields

\begin{mathletters}
\label{2.2}
\begin{eqnarray}
\frac{b_{n0}}{R} &=& \frac 3{4\pi}\,\frac{C(n0)}{1-\varsigma^3}\,
\left[\Lambda(R) - \varsigma^3\Lambda(a)\right] \label{2.2a} \\[1 ex]
\frac{b_{n2}}{R} &=& \frac 3{4\pi}\,\frac{C(n2)}{1-\varsigma^3}\,
\left[\Sigma(R) - \varsigma^3\Sigma(a)\right] \label{2.2b}
\end{eqnarray}
\end{mathletters}
with

\begin{mathletters}
\label{2.3}
\begin{eqnarray}
\Lambda(z) & \equiv & \frac{j_2(q_{n0}z)}{q_{n0}R} +
 D(n0)\,\frac{y_2(q_{n0}z)}{q_{n0}R}   \label{2.3a} \\[1 em]
\Sigma(z) & \equiv & \frac{j_2(q_{n2}z)}{q_{n2}R} -
3K(n2)\,\frac{j_2(k_{n2}z)}{k_{n2}R} +
D(n2)\,\frac{y_2(q_{n2}z)}{q_{n2}R} -
3\tilde D(n2)\,\frac{y_2(k_{n2}z)}{k_{n2}R}   \label{2.3b}
\end{eqnarray}
\end{mathletters}

The absorption {\it cross section\/}, defined as the ratio of the absorbed
energy to the incoming flux, can be calculated thanks to an {\it optical
theorem\/}, as proved e.g.\ by Weinberg \cite{wein-72}. According to that
theorem, the absorption cross section for a signal of frequency $\omega\/$
close to $\omega_N\/$, say, the frequency of the detector mode excited by
the incoming GW, is given by the expression

\begin{equation}
\sigma(\omega) = \frac{10\,\pi\eta c^2}{\omega^2}\,
		 \frac{\Gamma^2/4}{(\omega -\omega_N)^2 + \Gamma^2/4}
\label{2.4}
\end{equation}
where $\Gamma$ is the {\it linewitdh\/} of the mode ---which can be
arbitrarily small, as assumed in the previous section---, and $\eta\/$
is the dimensionless ratio

\begin{equation}
  \eta = \frac{\Gamma_{\rm grav}}{\Gamma} =
	 \frac{1}{\Gamma}\,\frac{P_{GW}}{E_{\rm osc}}
\label{2.5}
\end{equation}
where $P_{GW}$ is the energy {\it re-emitted\/} by the detector in the form
of GWs as a consequence of its being set to oscillate by the incoming signal.

It is now expedient to split up the energy emitted by the oscillating hollow
sphere into two pieces:

\begin{equation}
P_{GW} = P_{\rm pure\ tensor} + P_{\rm scalar-tensor}
\label{2.6}
\end{equation}
where $P_{\rm pure\ tensor}$ is given by General Relativity, and contains
only the usual $+$ and $\times$ amplitudes, while $P_{\rm scalar-tensor}$
is an added term which is only prdicted by Jordan-Brans-Dicke theory, and
has contributions from a scalar amplitude {\it plus\/} the $m\/$\,=\,0
quadrupole amplitude \cite{lobo-95,bian-98}. These terms are the following:

\begin{equation}
P_{\rm pure\ tensor} = \frac{2G\,\omega ^6}{5c^5}\,
\left[\left|Q_{kk}(\omega)\right|^2 -
\frac 13\,Q_{ij}^*(\omega)Q_{ij}(\omega)\right]
\label{2.7}
\end{equation}
and

\begin{equation}
P_{\rm scalar-tensor} = \frac{2G\,\omega ^6}{5c^5\,(2\Omega _{BD}+3)}\,
\left[\left|Q_{kk}(\omega)\right|^2 +
\frac 13\,Q_{ij}^*(\omega)Q_{ij}(\omega)\right]
\label{2.8}
\end{equation}
where $Q_{ij}(\omega)$ is the quadrupole moment of the hollow sphere:

\begin{equation}
Q_{ij}(\omega) = \int_{\rm Antenna}\,x_ix_j\,\varrho({\bf x},\omega)\,d^3x
\end{equation}
and $\Omega_{BD}\/$ is Brans--Dicke's parameter.

We shall omit any further reference to the pure tensor interaction, as it
was rather comprehensively discussed in reference \cite{bian-98}. We thus
concentrate in the sequel in the scalar tensor term.

\section{Scalar-tensor cross sections}

Explicit calculation shows that $P_{\rm scalar-tensor}$ is made up of two
contributions:

\begin{equation}
P_{\rm scalar-tensor} = P_{00} + P_{20}
\label{3.1}
\end{equation}
where $P_{00}$ is the scalar, or monopole contribution to the emitted power,
while $P_{20}$ comes from the central quadrupole mode which, as discussed in
\cite{bian-98} and \cite{lobo-95}, is excited together with monopole in JBD
theory. One must however recall that monopole and quadrupole modes of the
sphere happen at {\it different frequencies\/}, so that cross sections for
them only make sense if defined separately. More precisely,

\begin{mathletters}
\label{3.2}
\begin{eqnarray}
\sigma_{n0}(\omega) & = & \frac{10\pi\,\eta_{n0}\,c^2}{\omega^2}\,
  \frac{\Gamma_{n0}^2/4}{(\omega - \omega_{n0})^2 + \Gamma_{n0}^2/4}
  \label{3.2a} \\
\sigma_{n2}(\omega) & = & \frac{10\pi\,\eta_{n2}\,c^2}{\omega^2}\,
  \frac{\Gamma_{n2}^2/4}{(\omega - \omega_{n2})^2 + \Gamma_{n2}^2/4}
  \label{3.2b}
\end{eqnarray}
\end{mathletters}
where $\eta_{n0}$ and $\eta_{n2}$ are defined like in (\ref{2.5}), with all
terms referring to the corresponding modes. After some algebra one finds that

\begin{mathletters}
\label{3.3}
\begin{eqnarray}
\sigma_{n0}(\omega) & = & H_n\,\frac{GMv_S^2}{(\Omega_{BD}+2)\,c^3}\,
  \frac{\Gamma_{n0}^2/4}{(\omega - \omega_{n0})^2 + \Gamma_{n0}^2/4}
  \label{3.3a} \\
\sigma_{n2}(\omega) & = & F_n\,\frac{GMv_S^2}{(\Omega_{BD}+2)\,c^3}\,
  \frac{\Gamma_{n2}^2/4}{(\omega - \omega_{n2})^2 + \Gamma_{n2}^2/4}
  \label{3.3b}
\end{eqnarray}
\end{mathletters}

Here, we have defined the dimensionless quantities

\begin{mathletters}
\label{3.4}
\begin{eqnarray}
  H_n & = & \frac{4\pi^2}{ 9\,(1+\sigma_P)}\,(k_{n0}b_{n0})^2	\label{3.4a} \\
  F_n & = & \frac{8\pi^2}{15\,(1+\sigma_P)}\,(k_{n2}b_{n2})^2	\label{3.4b}
\end{eqnarray}
\end{mathletters}
where $\sigma_P\/$ represents the sphere material's Poisson ratio (most
often very close to a value of 1/3), and the $b_{nl}\/$ are defined in
(\ref{2.2}); $v_S\/$ is the speed of sound in the material of the sphere.

In tables \ref{t.1} and \ref{t.2} we give a few numerical values of the
above cross section coefficients.

\begin{table}
\caption{Eigenvalues $k_{n0}^R\/$, relative weights $D(n0)$ and $H_n\/$
coefficients for a hollow sphere with Poisson ratio $\sigma_P$\,=\,1/3.
Values are given for a few different thickness parameters $\varsigma$.}
\label{t.1}
\begin{tabular}{ddddd}
$\ \ \varsigma$ & $n$ & $\ \ k_{n0}^SR$ & \qquad\ \ $D(n0)$ & \qquad $H_n$ \\
\hline
0.01 & 1 & 5.48738 & -1.43328$\cdot 10^{-4}$ & 0.90929 \\
     & 1 & 12.2332 & -1.59636$\cdot 10^{-3}$ & 0.14194 \\
     & 2 & 18.6321 & -5.58961$\cdot 10^{-3}$ & 0.05926 \\
     & 4 & 24.9693 & -0.001279 		     & 0.03267 \\[1 ex]
0.10 & 1 & 5.45410 & -0.014218 & 0.89530 \\
     & 1 & 11.9241 & -0.151377 & 0.15048 \\
     & 2 & 17.7277 & -0.479543 & 0.04922 \\
     & 4 & 23.5416 & -0.859885 & 0.04311 \\[1 ex]
0.15 & 1 & 5.37709 & -0.045574 & 0.86076 \\
     & 2 & 11.3879 & -0.434591 & 0.17646 \\
     & 3 & 17.105  & -0.939629 & 0.05674 \\
     & 4 & 23.605  & -0.806574 & 0.05396 \\[1 ex]
0.25 & 1 & 5.04842 & -0.179999 & 0.73727 \\
     & 2 & 10.6515 & -0.960417 & 0.30532 \\
     & 3 & 17.8193 & -0.425087 & 0.04275 \\
     & 4 & 25.8063 &  0.440100 & 0.06347 \\[1 ex]
0.50 & 1 & 3.96914 & -0.631169 & 0.49429 \\
     & 2 & 13.2369 &  0.531684 & 0.58140 \\
     & 3 & 25.4531 &  0.245321 & 0.01728 \\
     & 4 & 37.9129 &  0.161117 & 0.07192 \\[1 ex]
0.75 & 1 & 3.26524 & -0.901244 & 0.43070 \\
     & 2 & 25.3468 &  0.188845 & 0.66284 \\
     & 3 & 50.3718 &  0.093173 & 0.00341 \\
     & 4 & 75.469  &  0.061981 & 0.07480 \\[1 ex]
0.90 & 1 & 2.98141 & -0.963552 & 0.42043 \\
     & 2 & 62.9027 &  0.067342 & 0.67689 \\
     & 3 & 125.699 &  0.033573 & 0.00047 \\
     & 4 & 188.519 &  0.022334 & 0.07538
\end{tabular}
\end{table}

\begin{table}
\caption{Eigenvalues $k_{n2}^R\/$, relative weights $K(n2)$, $D(n2)$,
$\tilde D(n2)$ and $F_n\/$ coefficients for a hollow sphere with Poisson
ratio $\sigma_P$\,=\,1/3. Values are given for a few different thickness
parameters $\varsigma$.}
\label{t.2}
\begin{tabular}{ddddddd}
$\ \ \varsigma$ & $n$ & $\ \ k_{n2}^SR$ & \qquad\ \ $K(n2)$ &
 \qquad\ \ $D(n2)$ &  \qquad\ \ $\tilde D(n2)$ & \qquad $F_n$ \\
\hline
0.10 & 1 &  2.63836 &  0.855799 & 0.000395 & -0.003142 & 2.94602 \\
     & 2 &  5.07358 &  0.751837 & 0.002351 & -0.018451 & 1.16934 \\
     & 3 & 10.96090 &  0.476073 & 0.009821 & -0.071685 & 0.02207 \\[1 ex]
0.15 & 1 &  2.61161 &  0.796019 &  0.001174 & -0.009288 & 2.86913 \\
     & 2 &  5.02815 &  0.723984 &  0.007028 & -0.053849 & 1.24153 \\
     & 3 &  8.25809 & -2.010150 & -0.094986 &  0.672786 & 0.08113 \\[1 ex]
0.25 & 1 &  2.49122 &  0.606536 &  0.003210 & -0.02494 & 2.55218 \\
     & 2 &  4.91223 &  0.647204 &  0.019483 & -0.13867 & 1.55022 \\
     & 3 &  8.24282 & -1.984426 & -0.126671 &  0.67506 & 0.05325 \\
     & 4 & 10.97725 &  0.432548 & -0.012194 &  0.02236 & 0.03503 \\[1 ex]
0.50 & 1 &  1.94340 &  0.300212 &  0.003041 & -0.02268 & 1.61978 \\
     & 2 &  5.06453 &  0.745258 &  0.005133 & -0.02889 & 2.29572 \\
     & 3 & 10.11189 &  1.795862 & -1.697480 &  2.98276 & 0.19707 \\
     & 4 & 15.91970 & -1.632550 & -1.965780 & -0.30953 & 0.17108 \\[1 ex]
0.75 & 1 &  1.44965 &  0.225040 &  0.001376 & -0.01017 & 1.15291 \\
     & 2 &  5.21599 &  0.910998 & -0.197532 &  0.40944 & 1.82276 \\
     & 3 & 13.93290 &  0.243382 &  0.748219 & -3.20130 & 1.08952 \\
     & 4 & 23.76319 &  0.550278 & -0.230203 & -0.81767 & 0.08114 \\[1 ex]
0.90 & 1 &  1.26565 &  0.213082 &  0.001019 & -0.00755 & 1.03864 \\
     & 2 &  4.97703 &  0.939420 & -0.323067 &  0.52279 & 1.54106 \\
     & 3 & 31.86429 &  6.012680 & -0.259533 &  4.05274 & 1.46486 \\
     & 4 & 61.29948 &  0.205362 & -0.673148 & -1.04369 & 0.13470
\end{tabular}
\end{table}

As already stressed in reference \cite{vega-98}, one of the main advantages
of a hollow sphere is that it enables to reach good sensitivities at lower
frequencies than a solid sphere. For example, a hollow sphere of the same
material and mass as a solid one ($\varsigma$\,=\,0) has eigenfrequencies
which are smaller by

\begin{equation}
 \omega_{nl}(\varsigma) = \omega_{nl}(\varsigma=0)\,(1-\varsigma^3)^{1/3}
 \label{3.5}
\end{equation}
for any mode indices $n\/$ and $l\/$. We now consider the detectability of
JBD GW waves coming from several interesting sources with a hollow sphere.

\section{Detectability of JBD signals}

The values of the coefficients $F_n\/$ and $H_n\/$, together with the
expressions (\ref{3.2}) for the cross sections of the hollow sphere, can
be used to estimate the detectability of typical GW signals.
In this section we report such estimates in terms of the maximum 
distances at which a coalescing compact
binary system and a gravitational collapse event give signal-to-noise
ratio equal to one in the
detectors. We consider these in turn.

\subsection{Binary systems}

We consider as a source of GWs a binary system formed by two neutron stars,
each of them with a mass of $m_1$\,=\,$m_2$\,=\,1.4\,$M_\odot$. In the inspiral
phase, the system emits a waveform of increasing amplitude and frequency 
(a "chirp") that can sweep up to the kHz range of frequency.
The {\it chirp mass\/}, which is the parameter that determines  
the frequency sweep rate of the chirp signal,
corresponding to this system is
$M_c\/$\,$\equiv$\,$(m_1m_2)^{3/5}\,(m_1+m_2)^{-1/5}$\,=\,1.22\,$M_\odot$,
and $\nu_{\rm [5\ cycles]}$\,=\,1270 Hz\footnote{
The frequency $\nu_{\rm [5\ cycles]}$ is the one the system has when it is
5 cycles away from coalescence. It is considered that beyond this frequency
disturbing effects distort the simple picture of a clean Newtonian binary 
system ---see \protect\cite{covi-96} for further references.}. Repeating 
the analysis carried on in  section five of \cite{brun-99} we find that the 
distance at which the signal-to-noise ratio, 
for a {\it quantum limited\/} detector, 
is equal to one is given by

\begin{mathletters}
\label{4.1}
\begin{eqnarray}
 r(\omega_{n0}) & = &\left[\frac{5\cdot 2^{1/3}}{32}\,
  \frac{1}{(\Omega_{BD}+2)(12\Omega_{BD}+19)}\,
  \frac{G^{5/3}M_c^{5/3}}{c^3}\,\frac{Mv_S^2}{\hbar\omega_{n0}^{4/3}}\,H_n
  \right]^{1/2}  \label{4.1a}  \\
 r(\omega_{n2}) & = &\left[\frac{5\cdot 2^{1/3}}{192}
  \frac{1}{(\Omega_{BD}+2)(12\Omega_{BD}+19)}\,
  \frac{G^{5/3}M_c^{5/3}}{c^3}\,\frac{Mv_S^2}{\hbar\omega_{n2}^{4/3}}\,F_n
  \right]^{1/2}  \label{4.1b}
\end{eqnarray}
\end{mathletters}

For a CuAL sphere, the speed of sound is $v_S\/$\,=\,4700 m/sec. We report
in table \ref{t.3} the maximum distances at which a JBD binary can be seen
with a 200 ton hollow spherical detector, including the size of the sphere
(diameter and thickness factor). The Brans-Dicke parameter $\Omega_{BD}\/$
has been given a value of 600. This high value has as a consequence that
only relatively nearby binaries can be scrutinised by means of their scalar
radiation of GWs. A slight improvement in sensitivity is appreciated as the
diameter increases in a fixed mass detector. 

\begin{table}
\caption{Eigenfrequencies, sizes and distances at which coalescing binaries
can be seen by monitoring of their emitted JBD GWs. Figures correspond to a
200 ton CuAl hollow sphere.}
\label{t.3}
\begin{tabular}{dddddd}
$\varsigma $ & $\Phi$ (m) & $\nu_{10}$(Hz) & $\nu_{12}$ (Hz)
& $r(\nu_{10})$ (kpc) & $r(\nu_{12})$ (kpc) \\ \hline
0.25 & 3.72 & 1243 & 613 & $92$ & 50 \\
0.50 & 3.88 & 937 & 459 & 91  & 48 \\
0.75 & 4.46  & 671  & 298 & 106 & 54   \\
0.90 & 5.74 & 476 & 202 & 131  & 67
\end{tabular}
\end{table}

\begin{table}
\caption{Eigenfrequencies, sizes and distances at which coalescing binaries
can be seen by monitoring of their emitted JBD GWs. Figures correspond to a
6 metres external diameter CuAl hollow sphere.}
\label{t.4}
\begin{tabular}{dddddd}
$\varsigma $ & $M$ (ton) & $\nu _{10}$(Hz) & $\nu_{12}$(Hz) &
$r(\nu_{10})$ (kpc) & $r(\nu_{12})$ (kpc) \\ \hline
0.25 & 832 & 770 & 380 & $257$  & 140  \\
0.50 & 740   & 609 & 296 & 233  & 124 \\
0.75 & 489  & 498 & 221 & 202 & 104   \\
0.90 & 230  & 455 & 193 & 145 & 74
\end{tabular}
\end{table}

\subsection{Gravitational collapse}

The signal associated to a gravitational collapse has recently been modeled,
within JBD theory, as a short pulse of amplitude $b\/$, whose value can be
estimated as \cite{bian-98}:

\begin{equation}
b \simeq 10^{-23}\,\left(\frac{500}{\omega_{BD}}\right)
  \left(\frac{M}{M_\odot}\right)\left(\frac{10\,Mpc}{r}\right)
\label{4.2}
\end{equation}

The minimum value of the Fourier transform of the amplitude of the scalar
wave, for a quantum limited detector at unit signal-to-noise ratio, is
given by

\begin{equation}
\left|b(\omega_{nl})\right|_{\min } = \left(
\frac{4\hbar}{Mv_S^2\omega_{nl}K_n}\right)^{1/2}
\label{4.3}
\end{equation}
where $K_n=2H_n$ for the mode with $l=0$ and $K_n=F_n/3$ for the 
mode with $l=2, m=0$.

The duration of the impulse, $\tau \approx 1/f_c$, is much shorter than
the decay time of the $nl\/$ mode, so that the relationship between $b\/$
and $b(\omega_{nl})$ is

\begin{equation}
b \approx \left| b(\omega_{nl})\right| f_c\qquad \text{at frequency\ \ \ }
  \omega_{nl} = 2\pi f_c
  \label{4.4}
\end{equation}
so that the minimum scalar wave amplitude detectable is

\begin{equation}
\left| b\right|_{\min}\approx\left(
  \frac{4\hbar\omega_{nl}}{\pi^2Mv_S^2K_n}\right)^{1/2}
  \label{4.5}
\end{equation}

Let us now consider a hollow sphere made of molibdenum, for which the speed
of sound is as high as $v_S\/$\,=\,5600 m/sec. For a given detector mass and
diameter, equation (\ref{4.5}) tells us which is the minimum signal detectable
with such detector. For example, a solid sphere of 31 tons and 1.8 metres in
diameter, is sensitive down to $b_{\min}$\,=\,1.5\,$\cdot$\,10$^{-22}$.
Equation (\ref{4.2}) can then be inverted to find which is the maximum
distance at which the source can be identified by the scalar waves it emits.
Taking a reasonable value of $\Omega_{BD}\/$\,=\,600, one finds that
$r(\nu_{10})$\,$\approx$\,0.6 Mpc.

Like before, we report in tables \ref{t.5} and \ref{t.6} the sensitivities
of the detector and consequent maximum distance at which the source appears
visible to the device for various values of the thickness parameter
$\varsigma\/$. In table~\ref{t.5} a constant detector mass of 31 tons has
been assumed for all thicknesses, and in table~\ref{t.6} a constant outer
diameter of 1.8 metres in all cases.

\begin{table}
\caption{Eigenfrequencies, maximum sensitivities and distances at which
a gravitational collapse can be seen by monitoring the scalar GWs it emits.
Figures correspond to a 31 ton Mb hollow sphere.}
\label{t.5}
\begin{tabular}{ddddd}
$\varsigma $ & $\phi$ (m) & $\nu_{10}$ (Hz) & $|b|_{\min }$ (10$^{-22}$)
& $r(\nu_{10})$ (Mpc) \\ \hline
0.00 & 1.80 & 3338 & 1.5  & 0.6  \\
0.25 & 1.82 & 3027 & 1.65 & 0.5  \\
0.50 & 1.88 & 2304 & 1.79 & 0.46 \\
0.75 & 2.16 & 1650 & 1.63 & 0.51 \\
0.90 & 2.78 & 1170 & 1.39 & 0.6
\end{tabular}
\end{table}

\begin{table}
\caption{Eigenfrequencies, maximum sensitivities and distances at which
a gravitational collapse can be seen by monitoring the scalar GWs it emits.
Figures correspond to a 1.8 metres outer diameter Mb hollow sphere.}
\label{t.6}
\begin{tabular}{ddddd}
$\varsigma $ & $M$ (ton) & $\nu_{10}$ (Hz) & $|b|_{\min }$ (10$^{-22}$)
& $r(\nu_{10})$ (Mpc) \\ \hline
0.00 & 31.0  & 3338 & 1.5  & 0.6  \\
0.25 & 30.52 & 3062 & 1.71 & 0.48 \\ 
0.50 & 27.12 & 2407 & 1.95 & 0.42 \\ 
0.75 & 17.92 & 1980 & 2.34 & 0.36 \\ 
0.90 & 8.4   & 1808 & 3.31 & 0.24
\end{tabular}
\end{table}

\section{Conclusions}

In this paper we have explored the ability of a hollow sphere to sense the
GWs emitted by radiating systems which obey the laws of Jordan-Brans-Dicke
theory rather than those of General Relativity. The difference between the
predictions of both theories is in the degrees of freedom of the signal:
while GR only predicts two polarisation states ---the usual $+$ and $\times$
amplitudes---, JBD predicts a scalar amplitude and one more quadrupole
amplitude ($m\/$\,=\,0) in addition to the $+$ and $\times$ amplitudes.
We have calculated the generic cross sections of the hollow spherical
detectors for the first few monopole and quadrupole harmonics, which are
relevant to the detection of JBD waves, then applied the results to the
analysis of how efficient the detection of such waves coming from a coalescing
compact binary system and a collapsing star, respectively, can be with a
hollow spherical GW detector.

In terms of the detector thickness it is seen that sensitivity appears to
be, in absolute figures, quite independent of the detector geometry, i.e.,
the weakest detectable signal in the best experimental circumstances, does
not change in order of magnitude over the entire range of possible thickness
parameter values. In other words, there is not much difference in sensitivity
between a hollow sphere and a massive sphere \cite{bian-98} for this kind of
detection problem. The one appreciable difference, however, is that one can
reach considerably lower frequencies in the sensitivity range of a hollow
sphere than can with a hollow sphere. Not surprisingly, the general facts
described in reference \cite{vega-98} also survive in this more general
framework of JBD theory.


\begin{thebibliography}{99}
\bibitem{will}  C.M.~Will, {\it Theory and Experiment in Gravitational  Physics}
(Cambridge University Press, Cambridge, 1993).

\bibitem{bian-98} M. Bianchi, M. Brunetti, E. Coccia, F. Fucito, and
J.A. Lobo, {\em Phys. Rev.} D 57, 4525 (1998).

\bibitem{lobo-95} J.A. Lobo, {\em Phys. Rev.} D 52, 591 (1995).

\bibitem{wp-75} R.V. Wagoner and H.J. Paik in {\em Proc. of the Int. Symposium
on Experimental Gravitation} (Accademia Nazionale dei Lincei, Rome, 1977).

\bibitem{vega-98} E. Coccia, V. Fafone, G. Frossati, J. A. Lobo, and J. A.
Ortega, {\em Phys. Rev.} D 57, 2051 (1998).

\bibitem{ligo} F.J. Raab (LIGO team), in E. Coccia, G.Pizzella, F.Ronga
(eds.), {\em Gravitational Wave Experiments}, Proceedings of the First
Edoardo Amaldi Conference, Frascati 1994 (World Scientific, Singapore, 1995).

\bibitem{virgo} A. Giazotto {\it et al.\/} (VIRGO collaboration), in
E. Coccia, G.Pizzella, F.Ronga (eds.), {\em Gravitational Wave Experiments},
Proceedings of the First Edoardo Amaldi Conference, Frascati 1994 (World
Scientific, Singapore, 1995).

\bibitem{jordan-59} P. Jordan, {\em Z. Phys.}, {\bf 157}, 112 (1959).

\bibitem{bd-61} C. Brans and R. H. Dicke, {\it Phys. Rev.\/} {\bf 124},
925 (1961).

\bibitem{love-44} A.E.H. Love, {\it A Treatise on the Mathematical Theory
of Elasticity}, Dover 1944.

\bibitem{landau-70} L. D. Landau and E. M. Lifshitz, {\it Theory of
Elasticity}, Pergamon 1970.

\bibitem{abram-72} M. Abramowitz and I.A. Stegun, {\it Handbook of
Mathematical Functions}, Dover 1972.

\bibitem{bian-96} M. Bianchi, E. Coccia, C. N. Colacino, V. Fafone, and
F. Fucito, {\em Class. and Quantum Grav.} {\bf 13}, 2865 (1996). 

\bibitem{wein-72} S. Weinberg, {\it Gravitation and Cosmology\/}, Wiley
\& sons, New York 1972.

\bibitem{covi-96} E. Coccia and V. Fafone, {\em Phys. Lett. A} {\bf 213},
16 (1996).

\bibitem{brun-99} M. Brunetti, E. Coccia, V. Fafone and F. Fucito, 
{\em Phys. Rev.} D 59, 044027 (1999).

\end{thebibliography}
\end{document}